\documentclass[english,aps,preprint]{revtex4}
\usepackage[T1]{fontenc}
\usepackage[latin1]{inputenc}
\usepackage{amssymb}

\makeatletter

\usepackage{babel}
\makeatother
\begin{document}

\title{The Semiclassical Coulomb Interaction}

\author{B. V. Carlson}

\email{brett@fis.ita.br}

\affiliation{Departamento de F\'{\i}sica, Instituto Tecnol\'{o}gico de Aeron\'{a}utica
- CTA, 12228-900 S\~{a}o Jos\'{e} dos Campos, Brazil}

\author{L.F. Canto}

\affiliation{Instituto de F\'{\i}sica, Universidade Federal do Rio de Janeiro,
21945-970 Rio de Janeiro, Brazil}

\author{M.S. Hussein}

\affiliation{Instituto de F\'{\i}sica, Universidade de Sao Paulo, 05389-970,
S\~{a}o Paulo, Brazil}

\begin{abstract}
The semiclassical Coulomb excitation interaction is at times expressed
in the Lorentz gauge in terms of the electromagnetic fields and a
contribution from the scalar electric potential. We point out that
the potential term can make spurious contributions to excitation cross
sections, especially when the decay of excited states is taken into
account. We show that, through an appropriate gauge transformation,
the excitation interaction can be expressed in terms of the electromagnetic
fields alone. 
\end{abstract}
\maketitle

\section*{Introduction}

Coulomb excitation has proven itself as an important tool for studying
the structure of both stable and exotic nuclei\cite{Ber88,Eml94,Cho95,Aum98}.
The phenomenon has been well-studied in a nonrelativistic context\cite{Ald65}
and studied perturbatively in a relativistic one\cite{Win79} by Alder
and Winther. More recent experimental studies have shown the need
to consider multiple Coulomb excitation at relativistic 
energies\cite{Eml94,Cho95,Aum98} and,
although the formalism of Ref. \onlinecite{Win79} can be extended
to permit such calculations, it is not easy to implement\cite{Bay99,Bay03a}.
An alternative semiclassical form of the Coulomb excitation interaction
was presented in Ref. \onlinecite{Ber96} and refined in a more
recent analysis\cite{Esb02}. There, the Coulomb excitation interaction
is expressed in the Lorentz gauge in terms of the electromagnetic
fields and the scalar electric potential. This mixed representation
of the interaction yields the results of Winther and Alder when used
perturbatively and is adequate for describing single and multiple
excitation of states of zero width at incident energies lower than
about 2 GeV/nucleon, where the results are almost identical to those
obtained by directly using the electromagnetic fields, as in a classical
treatment of the problem. At higher energies, coupled channel calculations
in the mixed and field representations yield results that are increasingly
discrepant, as the excitation cross section of the mixed representation
quickly grows to absurdly large values. Similar discrepancies have
also been observed in a recent comparison of relativistic Coulomb
excitation in the Lorentz and Coulomb gauges\cite{Bay05}.

Electromagnetic processes, such as Coulomb excitation, should be gauge-invariant.
Truncation of the coupled-channels model space can spoil this invariance
however\cite{Rum90}, permitting spurious, but gauge-removable terms,
such as the potential term in the Lorentz interaction, to contribute 
to the cross sections.  Baltz, Rhoades-Brown and Weneser have seen
similar large effects in their extensive study of $e^{+}e^{-}$ 
production\cite{Bal91,Bal93,Bal93b} and have found as well that, when 
performed with care, an appropriate gauge transformation
can greatly simplify calculations.

In the case of the mixed interaction of the Lorentz gauge, the effects
of the spurious potential term are exacerbated even further when the
excited states are allowed to decay. The mixed representation then
yields large, unphysical cross sections at all incident energies,
due to the loss of flux after excitation by the long-range potential
term. The production cross section of the decay products of a dipole
transition, in particular, grows as the square of the Lorentz factor
$\gamma$. This makes a treatment of multiple Coulomb excitation incorporating
fluctuation contributions, such as those of the Brink-Axel type
\cite{Car99a,Car99b,Car99c,Hus04},
unviable in the mixed representation, since finite widths are a fundamental
component of such models.

In the following, we first provide an estimate of the production cross
section of the decay products of a dipole transition induced by the
long-range potential term. We then demonstrate, through an appropriate
gauge transformation, that the Coulomb excitation component of the
interaction can indeed be expressed in terms of the electromagnetic
fields alone. Aside from making a satisfying parallel with the classical
case, the pure field representation of the excitation interaction,
known as the multipolar or Poincar\'e gauge\cite{Jac02}, provides physically
reasonable cross sections even when the excited states decay\cite{Car99c}.
A similar expression has been used in the treatment of Coulomb excitation
of plasmon resonances in metallic clusters\cite{Hus00}.

\section*{The Semiclassical Coulomb interaction }

In the usual semiclassical approximation to heavy-ion scattering,
the relative motion is described by a classical trajectory. The projectile-target
interaction is then a time-dependent function determined by this trajectory.
The semiclassical form of the electromagnetic interaction Hamiltonian
is given by\begin{equation}
V(t)=\int d^{3}x\,\left(\rho(\vec{x},t)\varphi(\vec{x},t)-\frac{1}{c}\vec{J}(\vec{x},t)\cdot\vec{A}(\vec{x},t)\right),\end{equation}
 where $\varphi(\vec{x},t)$ and $\vec{A}(\vec{x},t)$ are the scalar
and vector electromagnetic potentials due to the projectile, for which\begin{equation}
\vec{E}(\vec{x},t)=-\nabla\varphi(\vec{x},t)-\frac{1}{c}\frac{\partial\vec{A}(\vec{x},t)}{\partial t}\qquad\mbox{and}\qquad\vec{B}(\vec{x},t)=\nabla\times\vec{A}(\vec{x},t),\end{equation}
 and $\rho(\vec{x},t)$ and $\vec{J}(\vec{x},t)$ are the charge and
current density operators of the target nucleus. In high-energy collisions,
one usually uses the Liénard-Wiechert potential, which is the retarded
electromagnetic potential, in the Lorentz gauge, of a charged particle
moving on a straight line trajectory.

In Ref. \onlinecite{Ber96}, the Liénard-Wiechert potential
was used to obtain the Coulomb interaction for multipole excitation
of a target nucleus in a mixed representation that depends on both
the electric field and the scalar electric potential. The interaction
inducing transverse dipole excitations is written in terms of the
electric field alone. Due to the contribution of the vector potential,
the interaction inducing longitudinal transitions also includes a
potential-like term. It can be written as \begin{equation}
V_{1\parallel}(b,t)=V_{0}\left[\frac{\gamma vt}{\left(b^{2}+(\gamma vt)^{2}\right)^{3/2}}-e^{-i\omega t}\frac{\beta\gamma}{c}\frac{d\;}{dt}\left(\frac{e^{i\omega t}}{\left(b^{2}+(\gamma vt)^{2}\right)^{1/2}}\right)\right],\label{vpar}\end{equation}
 with $b$ being the impact parameter, $v$ the projectile velocity,
$\hbar\omega$ the excitation energy, $\beta=v/c$ and $\gamma$ the
associated Lorentz factor. The quantity $V_{0}$ represents the product
of the projectile charge, the matrix element for dipole excitation
and numerical factors. (Note that the factor $\mathcal{E}_{2}(\tau)$
defined in Eq. (26) of Ref. \onlinecite{Ber96} must be divided
by $\tau^{2}$ to provide the correct expression, which can be found
in Ref. \onlinecite{Bau00}.)

Let us now consider the time evolution of three states: the ground
state, the longitudinal dipole mode excited by Coulomb excitation
and the residual state fed by the decay of the latter. The time evolution
equations of the ground-state and dipole-mode amplitudes, $a_{0}(b,t)$
and $a_{1\parallel}(b,t)$, respectively, can be written in the interaction
picture as\cite{Car99c}\begin{eqnarray}
i\hbar\frac{d}{dt}a_{0}(b,t) & = & V_{1\parallel}(b,t)e^{-i\omega t}a_{1\parallel}(b,t)\nonumber\\
i\hbar\frac{d}{dt}a_{1\parallel}(b,t) & = & V_{1\parallel}(b,t)e^{i\omega t}a_{0}(b,t)-i\frac{\Gamma}{2}\, a_{1\parallel}(b,t),\end{eqnarray}
 where $\Gamma$ is the width of the dipole resonance. The residual state
is fed incoherently by the decay of the dipole mode. The time evolution
equation for its occupation probability can be written as \begin{equation}
\frac{dP_{dec}}{dt}(b,t)=\frac{\Gamma}{\hbar}\left|a_{1\parallel}(b,t)\right|^{2}.\end{equation}
 Coupled coherent-incoherent evolution equations such as these can
be consistently formulated in terms of the density matrix, as shown
in Refs. \onlinecite{Car99c} and \onlinecite{Hus04}. The formulation above is
sufficient for our purposes here.

Since the interaction tends to zero as $b\rightarrow\infty$, first
order perturbation theory is valid at large values of the impact parameter,
where depletion of the ground-state and occupation of the longitudinal
excitation can be neglected on the right side of the equation. We
can then approximate the amplitude for excitation of the longitudinal
mode by retaining only the second term in $V_{1\parallel}(b,t)$, which
decreases as $b^{-1}$, as
 \begin{eqnarray}
a_{1\parallel}(b,t) & \approx & -\frac{i}{\hbar}\int_{-\infty}^{t}ds\, V_{1\parallel}(b,s)e^{i\omega s}\\
 & = & iV_{0}\frac{\beta\gamma}{\hbar c}\frac{e^{i\omega t}}{\left(b^{2}+(\gamma vt)^{2}\right)^{1/2}}+{\mathcal{O}}\left(\frac{1}{b^{3}}\right)\,.\nonumber \end{eqnarray}
 It is clear that the contribution of the $1/b$ term vanishes as
$t\rightarrow\infty$ and that it thus makes no net contribution to
the excitation, when no flux is absorbed from the excited mode and
ground-state depletion is negligible. These are the assumptions used
in the usual perturbative calculation of the asymptotic amplitude.
When the dipole mode decays, however, this term does contribute to
the occupation of the residual state, with the asymptotic occupation
probability of that state being\begin{eqnarray}
P_{dec}(b) & = & \frac{\Gamma}{\hbar}\int_{-\infty}^{\infty}dt\,\left|a_{1\parallel}(b,t)\right|^{2}\\
 & \approx & \frac{\beta\gamma}{b}\frac{\Gamma V_{0}^{2}}{\left(\hbar c\right)^{3}}\,.\nonumber \end{eqnarray}
 Using this perturbative probability, we can estimate the cross section
for production of the decay product to be at least \begin{equation}
\sigma_{dec}\approx2\pi\int_{b_{min}}^{b_{max}}P_{dec}(b)\, bdb=2\pi\beta\gamma\frac{\Gamma V_{0}^{2}}{\left(\hbar c\right)^{3}}\left(b_{max}-b_{min}\right),\end{equation}
 where $b_{min}$ is the minimum value of the impact parameter at
which the perturbative approximation is valid and $b_{max}=\gamma v/\omega_{0}$
is determined by limiting the impact parameter to values for which
the adiabaticity parameter $\xi=\frac{\omega_{0}b}{\gamma v}$ is
less than one\cite{Win79}, with $\hbar\omega_{0}$ the minimum value
of the excitation energy considered as contributing to the dipole
mode. The resulting cross section thus grows with energy as $\gamma^2$.

The absurd result furnished by the above estimate is confirmed by
the full coupled-channels calculations given in the table. The calculations
were performed for the system $^{208}$Pb + $^{208}$Pb using the
three channels described above, with a dipole excitation energy of
$\hbar\omega=13.4$ MeV, a reduced matrix element in accord with the
giant dipole resonance sum rule and a minimum excitation energy of
$\hbar\omega_{0}=8$ MeV. We have labeled the cross sections of the
table as dipole-mode excitation cross sections. In the case of zero-width,
they are in fact the cross sections for excitation of this mode. In
the case of finite width, the values represent the flux that was excited
to the dipole mode to later decay to the residual state. 

Similar calculations were reported in Ref. \cite{Bau00}.
Comparison of their zero-width results with ours and with
those of Ref. \cite{Bay03b} leads us to conclude that they performed
the calculations with the pure field form that we are advocating. 
The trend of their finite-width
results also leads us to infer that they took the width into account
in the decay of the dipole mode but not in its excitation.

\begin{table}

\caption{Dipole mode excitation cross section of $^{208}$Pb incident on $^{208}$Pb
using the pure field and the mixed (Lorentz) representations of the
interaction.}
\begin{ruledtabular}
\begin{tabular}{|c|c|c|c|}
\hline 
Representation &
 $\Gamma$(MeV) &
 $\sigma$(b) at 1 GeV/nucleon &
 $\sigma$(b) at 10 GeV/nucleon\tabularnewline
\hline
field&
 0&
 4.26&
 13.98\tabularnewline
\hline
field&
 4&
 4.53&
 13.73\tabularnewline
\hline
mixed&
 0&
 4.46&
 98.84\tabularnewline
\hline
mixed&
 4&
 15.09&
 334.06 \tabularnewline
\hline
\end{tabular}
\end{ruledtabular}
\end{table}

We thus conclude that the spurious potential term in the interaction
produces absurdly large cross sections when decay of the excited state
is taken into account. Even when the width of the state is zero, this
term, abetted by depletion of the ground-state, introduces contributions
that increase with incident energy and also lead to absurdly large
cross sections, such as those at 10 GeV/nucleon shown in the last column of
the table. The unphysical results in both cases are attributable to
the unphysical $1/b$ term in the truncated coupled-channels calculations.
We can also argue against such a term on simple physical grounds:
we expect the polarization that produces the excitation of the target
nucleus to depend on the gradient of the potential, that is the electric
field, rather than the potential itself. In the following, we will
show how the electromagnetic interaction Hamiltonian can be recast
in a more physical form.

\section*{Expansion of the interaction Hamiltonian}

We want to obtain the first few terms contributing to the interaction
in the expansion of the electromagnetic fields about the center of
the target nucleus, $\vec{x}=0$. Such an expansion is reasonable
if the fields are slowly varying over the extent of the nucleus.

We thus take\begin{equation}
\int d^{3}x\,\rho(\vec{x},t)\varphi(\vec{x},t)\approx\int d^{3}x\,\rho(\vec{x},t)\left(\varphi_{0}(t)+\vec{x}\cdot\nabla\varphi_{0}(t)+\frac{1}{2}\vec{x}\vec{x}\cdot\nabla\nabla\varphi_{0}(t)+\cdots\right)\end{equation}
 and\begin{equation}
\int d^{3}x\,\vec{J}(\vec{x},t)\cdot\vec{A}(\vec{x},t)\approx\int d^{3}x\,\vec{J}(\vec{x},t)\cdot\left(\vec{A}_{0}(t)+\vec{x}\cdot\nabla\vec{A}_{0}(t)+\cdots\right),\end{equation}
 where the subscript $0$ on the fields and their derivatives denotes
the evaluation of these quantities at the point $\vec{x}=0$.

Evaluation of the scalar potential terms is straightforward. Evaluation
of the vector potential terms requires a bit more work. We use the
continuity equation, \begin{equation}
\nabla\cdot\vec{J}+\frac{\partial\rho}{\partial t}=0,\label{conteq}\end{equation}
 to obtain two supplementary identities\cite{Jac75}:\begin{equation}
\int d^{3}x\, J_{k}=\int d^{3}x\,\left(\nabla\cdot\left(x_{k}\vec{J}\right)-x_{k}\nabla\cdot\vec{J}\right)=\int d^{3}x\, x_{k}\frac{\partial\rho}{\partial t},\label{j0}\end{equation}
 where the integral of the exact divergence is zero due to the finite
extent of $\vec{J}$, and, after a similar calculation,\begin{equation}
\int d^{3}x\,\left(J_{k}x_{i}+J_{i}x_{k}\right)=\int d^{3}x\, x_{k}x_{i}\frac{\partial\rho}{\partial t}.\label{j1}\end{equation}
 Using the first of these, we can write \begin{equation}
\int d^{3}x\,\vec{J}(\vec{x},t)\cdot\vec{A}_{0}(t)=\sum_{k}\int d^{3}x\, J_{k}(\vec{x},t)A_{0k}(t)=\sum_{k}\int d^{3}x\,\frac{\partial\rho}{\partial t}x_{k}A_{0k}(t)=\int d^{3}x\,\frac{\partial\rho}{\partial t}\vec{x}\cdot\vec{A}_{0}(t).\end{equation}
 Using the second, we find, with a bit more work,\begin{equation}
\int d^{3}x\,\vec{J}(\vec{x},t)\cdot(\vec{x}\cdot\nabla)\vec{A}_{0}(t)=\frac{1}{2}\int d^{3}x\,\frac{\partial\rho}{\partial t}\vec{x}\cdot(\vec{x}\cdot\nabla)\vec{A}_{0}(t)+\frac{1}{2}\int d^{3}x\,\left(\vec{x}\times\vec{J}\right)\cdot\left(\nabla\times\vec{A}_{0}(t)\right).\end{equation}

Putting all the pieces together, we have\begin{eqnarray}
V(t) & = & \int d^{3}x\,\left(\rho(\vec{x},t)\varphi(\vec{x},t)-\frac{1}{c}\vec{J}(\vec{x},t)\cdot\vec{A}(\vec{x},t)\right)\label{vv}\\
 & = & \int d^{3}x\,\rho(\vec{x},t)\left(\varphi_{0}(t)+\vec{x}\cdot\nabla\varphi_{0}(t)+\frac{1}{2}\vec{x}\vec{x}\cdot\nabla\nabla\varphi_{0}(t)+\dots\right)\nonumber \\
 &  & -\frac{1}{c}\int d^{3}x\,\frac{\partial\rho}{\partial t}\vec{x}\cdot\left(\vec{A}_{0}(t)+\frac{1}{2}(\vec{x}\cdot\nabla)\vec{A}_{0}(t)+\dots\right)\nonumber \\
 &  & -\frac{1}{2c}\int d^{3}x\,\left(\vec{x}\times\vec{J}\right)\cdot\left(\nabla\times\vec{A}_{0}(t)\right)+\dots\nonumber \end{eqnarray}
 The unphysical long-range contribution to the excitation interaction
of Ref. \onlinecite{Ber96} can be traced to the term containing
the time derivative of the charge density, $\partial\rho/\partial t$.

\section*{The Gauge transformation}

We would now like to eliminate the terms containing the time derivative
of the charge density, $\partial\rho/\partial t$, in Eq.(\ref{vv}).
We can do this by making the gauge transformation $\Lambda(\vec{x},t)$
given by the factor multiplying this term, \begin{equation}
\Lambda(\vec{x},t)=-\vec{x}\cdot\left(\vec{A}_{0}(t)+\frac{1}{2}(\vec{x}\cdot\nabla)\vec{A}_{0}(t)+\dots\right)=-\int_{0}^{1}du\,\vec{x}\cdot\vec{A}(u\vec{x},t).\label{gauge}\end{equation}
 The vector potential that results from the gauge transformation can
be expanded as \begin{eqnarray}
\vec{A}^{\prime}(\vec{x},t) & = & \vec{A}(\vec{x},t)+\nabla\Lambda(\vec{x},t)\\
 & = & -\frac{1}{2}\vec{x}\times\left(\nabla\times\vec{A}_{0}\right)+\dots,\nonumber \end{eqnarray}
 while the transformed scalar potential can be expanded as \begin{eqnarray}
\varphi^{\prime}(\vec{x},t) & = & \varphi(\vec{x},t)-\frac{1}{c}\frac{\partial\Lambda}{\partial t}\\
 & = & \varphi_{0}(t)+\vec{x}\cdot\left(\nabla\varphi_{0}(t)+\frac{1}{c}\frac{\partial\vec{A}_{0}}{\partial t}\right)+\frac{1}{2}\vec{x}\vec{x}\cdot\nabla\left(\nabla\varphi_{0}(t)+\frac{1}{c}\frac{\partial\vec{A}_{0}}{\partial t}\right)+\cdots.\nonumber \end{eqnarray}

We can then rewrite the interaction as,\begin{eqnarray}
V(t) & \approx & \int d^{3}x\,\rho(\vec{x},t)\left(\varphi_{0}(t)+\vec{x}\cdot\left(\nabla\varphi_{0}(t)+\frac{1}{c}\frac{\partial\vec{A}_{0}}{\partial t}\right)+\frac{1}{2}\vec{x}\vec{x}\cdot\nabla\left(\nabla\varphi_{0}(t)+\frac{1}{c}\frac{\partial\vec{A}_{0}}{\partial t}\right)+\cdots\right)\nonumber \\
 &  & -\frac{1}{2c}\int d^{3}x\,\left(\vec{x}\times\vec{J}\right)\cdot\left(\nabla\times\vec{A}_{0}(t)\right)+\cdots,\end{eqnarray}
 which we can express in terms of the electromagnetic fields $\vec{E}_{0}$
and $\vec{B}_{0}$ at $\vec{x}=0$ as\begin{eqnarray}
V(t) & \approx & \int d^{3}x\,\rho(\vec{x},t)\left(\varphi_{0}(t)-\vec{x}\cdot\vec{E}_{0}(t)-\frac{1}{2}\vec{x}\vec{x}\cdot\nabla\vec{E}_{0}(t)+\cdots\right)\nonumber \\
 &  & \qquad-\frac{1}{2c}\int d^{3}x\,\left(\vec{x}\times\vec{J}\right)\cdot\vec{B}_{0}(t)+\cdots.\end{eqnarray}
 If we assume as well that the field-producing charge does not overlap
with the nuclear one, we can write the electric quadrupole term as
\begin{equation}
\vec{x}\vec{x}\cdot\nabla\vec{E}_{0}=\sum_{i,j}\, x_{i}x_{j}\partial_{j}E_{0i}=\sum_{i,j}\,(x_{i}x_{j}-\vec{x}^{2}\delta_{ij}/3)\partial_{j}E_{0i},\end{equation}
 since, in that case, $\nabla\cdot\vec{E}_{0}=0$.

We can then write the interaction in a form that parallels the classical
(Cartesian) multipole expansion\cite{Jac75}, as\begin{equation}
V(t)=q\varphi(t)-\vec{d}\cdot\vec{E}_{0}(t)-\frac{1}{2}\sum_{i,j}Q_{ij}\partial_{j}E_{0i}(t)-\vec{m}\cdot\vec{B}_{0}(t)+\cdots,\end{equation}
 where $q$ is the charge,\begin{equation}
q=\int d^{3}x\,\rho(\vec{x},t),\end{equation}
 $\vec{d}$ is the electric dipole operator,\begin{equation}
\vec{d}=\int d^{3}x\,\vec{x}\,\rho(\vec{x},t),\end{equation}
 the $Q_{ij}$ are the traceless electric quadrupole operators,\begin{equation}
Q_{ij}=\int d^{3}x\,(x_{i}x_{j}-\vec{x}^{2}\delta_{ij}/3)\,\rho(\vec{x},t),\end{equation}
 and $\vec{m}$ is the magnetic dipole operator,\begin{equation}
\vec{m}=\frac{1}{2c}\int d^{3}x\,\vec{x}\times\vec{J}(\vec{x},t).\end{equation}
 The multipole expansion does not depend on whether we have included
magnetization currents, exchange contributions or other corrections
that determine the detailed structure of the target charge and current
densities. The only property of these densities that we have used
is the continuity equation, Eq. (\ref{conteq}), which should be valid
in any case.

The multipole expansion given here can be extended to all orders without
great difficulty. It can be expressed compactly as \begin{equation}
V(t)=q\varphi(t)-\int d^{3}x\,\rho(\vec{x},t)\,\vec{x}\cdot\int_{0}^{1}du\,\vec{E}(u\vec{x},t)-\frac{1}{c}\int d^{3}x\,\left(\vec{x}\times\vec{J}\right)\cdot\int_{0}^{1}duu\,\vec{B}(u\vec{x},t).\end{equation}
 It satisfies the gauge condition $\vec{x}\cdot\vec{A}(\vec{x},t)=0$,
which can be rewritten as \begin{equation}
\int_{0}^{1}du\,\vec{x}\cdot\vec{A}(u\vec{x},t)=0,\end{equation}
 (see Eq.{[}\ref{gauge}{]}) and is known as the multipolar or Poincar\'e
gauge\cite{Jac02}. A drawback to our formulation is that it has been
performed in a Cartesian rather than a spherical basis. However, the
expansion in the spherical basis should be directly deducible from
the Cartesian one, since the two are equivalent.

\section*{Discussion}

In the physically intuitive form given above, the electromagnetic
interaction poses no problem, even when the excitation and decay of
resonant states is considered. As an example, we take the Liénard-Wiechert
potential due to a relativistic nucleus of charge $Z$ passing on
a straight-line trajectory with velocity $v$ in the $\hat{z}$ direction
at a distance $b_{0}$ from the center of the charge/current distribution
of the target. We then have\cite{Jac75}\begin{equation}
\varphi(\vec{b},z,t)=\gamma\frac{Ze}{\sqrt{(\vec{b}-\vec{b}_{0})^{2}+\gamma^{2}(z-vt)^{2}}}\qquad\mbox{and}\qquad\vec{A}(\vec{b},z,t)=\frac{v}{c}\hat{z}\,\varphi(\vec{b},z,t).\end{equation}
 The transverse and longitudinal components of the electric field,
which induce dipole excitation, are \begin{equation}
\vec{E}_{0\perp}(t)=-\gamma\vec{b}_{0}\frac{Ze}{\left(\vec{b}_{0}^{2}+(\gamma vt)^{2}\right)^{3/2}}\qquad\mbox{and}\qquad E_{0||}(t)=-\gamma vt\frac{Ze}{\left(\vec{b}_{0}^{2}+(\gamma vt)^{2}\right)^{3/2}}\,,\end{equation}
 and tend to zero as $b_{0}^{-2}$and $b_{0}^{-3}$, respectively,
as the impact parameter increases. At high energies, the excitation
is dominated by the transverse modes, which produce a cross section
that grows as $\ln(b_{max}/b_{min})$, as expected\cite{Win79}.

Bayman and Zardi\cite{Bay03b} have observed that the mixed representation
of the interaction given in Ref. \onlinecite{Ber96} neglects
relativistic corrections to the quadrupole and higher multipole terms,
which become important at high energies. These are included and discussed
in their work and in Ref. \onlinecite{Esb02}. They are also
taken into account correctly (and automatically) when the pure field
representation is used.

As mentioned in the introduction, Bayman and Zardi have also recently
compared calculations of relativistic Coulomb excitation in the Lorentz
and Coulomb gauges\cite{Bay05}. They find that the cross sections
in the Lorentz gauge increases dramatically with respect to those
in the Coulomb gauge at energies above about 2 GeV/nucleon. Although
the Coulomb gauge is not equivalent to the one we present here, we
suspect that it too might include only physical contributions to coupled-channels
cross sections. We are of the opinion, however, that the optimal form
of the interaction in such calculations is the one given here,
in which the excitation interaction is expressed in terms of the physical
fields and their derivatives.

\section*{Conclusion}

We have shown that, through an appropriate gauge transformation, the
Coulomb excitation component of the semiclassical form of the electromagnetic
interaction can be expressed in terms of the electromagnetic fields
and their derivatives. Aside from making a satisfying parallel with
the classical case, the pure field representation of the excitation
interaction, known as the multipolar or Poincar\'e gauge, provides physically
reasonable cross sections, even when resonance excitation and decay
is taken into account.

\section*{Acknowledgments}

B.V.C. and M.S.H. acknowledge the support of the Funda\c{c}\~{a}o
de Amparo a Pesquisa do Estado de S\~{a}o Paulo, FAPESP. L.F.C.,
B.V.C. and M.S.H. acknowledge support from the Conselho Nacional de
Pesquisa e Desenvolvimento, CNPq.


\begin{thebibliography}{10}
\bibitem{Ber88}C. A. Bertulani and G. Baur, Phys. Rep. \textbf{163}, 299 (1988). 
\bibitem{Eml94}H. Emling, Progr. Part.Nucl. Phys. \textbf{33}, 792 (1994). 
\bibitem{Cho95}Ph. Chomaz and N. Frascaria, Phys. Rep. \textbf{252}, 275 (1995). 
\bibitem{Aum98}T. Aumann, P.F. Bortignon and H. Emling, Ann. Rev. Nucl. Sci \textbf{48},
351 (1998). 
\bibitem{Ald65}K. Alder and A. Winther, \textit{Coulomb Excitation}, (Academic Press,
New York, 1965). 
\bibitem{Win79}A. Winther and K. Alder, Nucl. Phys. A \textbf{319}, 518 (1979). 
\bibitem{Bay99}B.F. Bayman and F. Zardi, Phys. Rev. C \textbf{59}, 2189, (1999). 
\bibitem{Bay03a}B. F. Bayman and F. Zardi, Phys. Rev. C \textbf{68}, 014905 (2003). 
\bibitem{Ber96}C. A. Bertulani, L. F. Canto, M. S. Hussein and A. F. R. de Toledo
Piza, Phys. Rev. C \textbf{53}, 334 (1997). 
\bibitem{Esb02}H. Esbensen and C. A. Bertulani, Phys. Rev. C \textbf{65}, 024605
(2002). 
\bibitem{Bay05}B. F. Bayman and F. Zardi, Phys. Rev. C \textbf{71}, 014904 (2005). 
\bibitem{Rum90}K. Rumrich, W. Greiner, and G. Soff, Phys. Lett. A \textbf{149}, 17 (1990). 
\bibitem{Bal91}A.J. Baltz, M.J. Rhoades-Brown, and J. Weneser, Phys. Rev. A \textbf{44},
5569 (1991). 
\bibitem{Bal93}A.J. Baltz, M.J. Rhoades-Brown, and J. Weneser, Phys. Rev. A \textbf{47},
3444 (1993). 
\bibitem{Bal93b}A.J. Baltz, M.J. Rhoades-Brown, and J. Weneser, Phys. Rev. A \textbf{48},
2002 (1993). 
\bibitem{Car99a}B.V. Carlson, L.F. Canto, S. Cruz-Barrios, M.S. Hussein, A.F.R. de
Toledo Piza, Phys. Rev. C \textbf{59}, 2689 (1999). 
\bibitem{Car99b}B.V. Carlson, L.F. Canto, S. Cruz-Barrios, M.S. Hussein, A.F.R. de
Toledo Piza, Ann. Phys.(N.Y.) \textbf{276}, 111 (1999). 
\bibitem{Car99c}B.V. Carlson, M.S. Hussein, A.F.R. de Toledo Piza, L.F. Canto, Phys.
Rev. C \textbf{60}, 014604 (1999). 
\bibitem{Hus04}M.S. Hussein, B.V. Carlson and L.F. Canto,
  Nucl. Phys. A \textbf{731}, 163 (2004). 
\bibitem{Jac02}J.D. Jackson, Am. J. Phys. \textbf{70}, 917 (2002). 
\bibitem{Hus00}M. S. Hussein, V. Kharchenko, L. F. Canto, and R. Donangelo, Ann. Phys.
(N.Y.) \textbf{284}, 178 (2000). 
\bibitem{Bau00}G. Baur, C.A. Bertulani and D. Dolci, Eur. Phys. J. A \textbf{7}, 55
(2000). 
\bibitem{Jac75}J. D. Jackson, \textit{Classical Electrodynamics}, (Wiley, New York,
1975). 
\bibitem{Bay03b}B. F. Bayman and F. Zardi, Phys. Rev. C \textbf{67},
  017901 (2003). 
\end{thebibliography}
\end{document}